\documentclass[twocolumn,showpacs,preprintnumbers,amsmath,amssymb,aps,prb]{revtex4}

\usepackage{graphics}
\usepackage{color}
\begin{document}

\title{Exchange Constants and Neutron Spectra of Iron Pnictide Materials}

\author{Rajiv R.P. Singh}
\affiliation{University of California Davis, CA 95616, USA}

\date{\rm\today}

\begin{abstract}
We show that the neutron scattering spectra observed in CaFe$_2$As$_2$
by Zhao et al (arXiv:0903.2686v1)
and the highly unusual spatially anisotropic 
exchange constants in the iron pnictides
derived by Han et al (Phys. Rev. Lett. 102, 107003 (2009)), using electronic
structure calculations, can be explained by assuming a role for orbital
order in these materials. We write down a simple model Hamiltonian
with tertagonal symmetry, whose spin-wave spectra describes the
observed dispersion relations. We further argue that
these materials have orbital selective Mott transition, which
is driven by superexchange between neighboring iron atoms.
We suggest that reduced spin and quasi-one dimensionality
and not frustration are responsible for the reduced moments in
these materials.

\end{abstract}


\maketitle

The discovery of superconductivity at relatively high temperatures
in the iron pnictide family of materials has invigorated condensed
matter research.\cite{kamihara} In order to develop a fundamental understanding of
high temperature superconductivity, it is important to understand the
relationship and differences between these pnictides and the even
higher temperature superconductors, the cuprates. Both family of
materials are layered quasi-two dimensional electronic systems,
which at stochiometry develop antiferromagnetic order. Upon electronic
doping of these layers, they become superconducting. One of the
key differences between the two is that the cuprates, at stochiometry,
are in many respects classic Mott insulators. In local density approximation (LDA)
based calculations they are found to be metallic and only strong
correlation effects drive them to insulating and antiferromagnetic
behavior. In contrast, the parent state of iron pnictide materials
is a metal, whose conductivity increases upon the onset of
antiferromagnetism.\cite{cruz} This metallic antiferromagnetic state is
also reproduced in the LDA calculations.\cite{dong,yin,cao,ma,yildrim,wu,haule} 

Furthermore, the cuprate ions have a single d-hole, 
whereas the iron ions in the pnictides have multiple d electrons. 
Thus the former have spin-half, 
making up a system that has maximum quantum fluctuations
and possibly exotic incipient quantum spin-liquid behavior,\cite{anderson}
whereas the local moment in the latter could be much larger and thus
the magnetism more conventional.
Any role of strong correlations in the pnictides
has also been questioned, and this would have a strong bearing on
any universal understanding of the mechanism of superconductivity
in these family of materials.

We focus here on the parent insulating material. Magnetism in these
materials has been highly debated.\cite{wu,haule,si,fang,xu,mazin,raghu,ran} 
Some argue that these materials
are spin-density-waves driven by nesting of fermi-surfaces and they
should be viewd as weakly correlated itinerant magnets. Others
have argued that despite the metallic behavior,
strong correlations are present and the correct starting point for understanding the
magnetism in these materials is a Heisenberg type model. Indeed, 
neutron scattering spectra can we be well understood by the latter approach.

The conventional strong correlation view has been that the magnetism
is highly frustrated by having nearest neighbor antiferromagnetic 
exchange $J_1$ compete with second neighbor antiferromagnetic 
exchange $J_2$.\cite{haule,si,fang,xu}
It is well known that this leads to a ($\pi,0$) ordered state at large $J_2$ as
observed in experiments. In this case, there is an additional Ising
degree of freedom associated with the direction of antiferromagnetic and
ferromagnetic neighbors, which leads to a finite temperature phase
transition.\cite{chandra,capriotti} 
At, or below this transition, small anisotropies and or
weak three-dimensional coupling should lead to antiferromagnetic long-range order.
The Ising transition should couple to lattice distortions and lead to
a loss of tetragonal symmetry. Indeed, such structural and magnetic transitions
are observed either at the same temperature, or with the structural transition being
slightly higher in temperature, as expected from theory.
The proximity to a quantum critical point near
$J_2=J_1/2$ can lead to a highly reduced moment which fluctuates with
small changes in material parameters.\cite{uhrig}

The key to distinguishing different scenarios is Neutron Scattering
and the measurement of spin-wave dispersion in the full 
Brillouin zone.\cite{zhao1,ewings}
In this respect, the $J_1-J_2$ model scenario is also of deep theoretical
significance from the point of view of quantum magnetism, because the 
($\pi,0$) order is selected by an order by disorder phenomena.\cite{shender} 
The accidental degeneracy of the classical model leads to zero energy
states near $4$-points of the Brillouin zone ($0,0$), ($\pi,0$), ($0,\pi$),
($\pi,\pi$). Quantum fluctuations, which lift the accedental degeneracy,
cause a gap at ($0,\pi$) and ($\pi,\pi$), leaving gapless excitations
only at ($0,0$) and ($\pi,0$) as required by Goldstone's theorem.\cite{uhrig,singh}

Recently Zhao et al have performed\cite{zhao} neutron scattering experiments
on the material CaFe$_2$As$_2$. The experiments show sharply defined
spin waves throught the zone, not just at very low energies. The
absence of a Stoner decay, leads them to strongly favor a local
moment type picture for the system. However, a big surprise is
the finding that the spin-wave energy along the $K$ direction is 
a maximum at ($\pi,\pi$). We will show below that this requires
that the nearest neighbor exchange be strong and antiferromagnetic in
one direction and weak and ferromagnetic in the other. Quantum
fluctuations can lead to a gap at ($\pi,\pi$) but they cannot turn it
into a maximum. The materials clearly lack tetragonal symmetry and
so such a result can be considered as resulting from lack of
tetragonal symmetry. However, one needs to understand why
a small distortion can lead to such a dramatic difference in exchange
constants. Indeed,
this is exactly what the elctronic structure calculations of
Han et al using local spin-density approximation (LSDA) find, within the tetragonal phase.\cite{yin,han} 
These results have not yet found a simple explanation. The main purpose
of this paper is to give a simple scenario by which this happens.

Rather than trying to write down a complete Hamiltonian for the low energy behavior
of this system, we proceed in a step by step manner to uncover the
physics behind the unusual neutron scattering spectra and the
unusual LSDA results.
We begin with a two orbital model (which we take as $d_{xz}$ and $d_{yz}$)
on the square-lattice in the XY plane,
and assume that the total occupancy of these orbitals is one. Clues to
the origin of such a model is discussed later.
When there are occupied $d_{xz}$ orbitals on neighboring sites along the x-axis,
there is a strong antiferromagnetic super-exchange $J_1$ between
the orbitals. Similarly, if there are $d_{yz}$ orbitals occupied on
neighboring sites along the y-axis, there is strong antiferromagnetic
exchange $J_1$. One can visualize that this is a direct exchange
between the d-orbitals and is only non-zero when the orbitals are
oriented towards each other. In addition, we assume there is a ferromagnetic
nearest neighbor exchange $J_f$ and an antiferromagnetic second
neighbor exchange $J_2$. One can imagine that both $J_f$ and $J_2$ are
mediated by the As atoms and are much weaker than $J_1$. The precise
variation of $J_2$ and $J_f$ with orbital occupation is not
important in our model. In general,
$J_2$ and magnetic frustration does not play a significant role in
our considerations. So, the
full Hamiltonian, with tertagonal symmetry, can be written as:

\begin{eqnarray}
H =&J_1 \sum_i [ \vec S_i\cdot \vec S_{\i+\hat x} P^x_i P^x_{i+\hat x}
              + \vec S_i\cdot \vec S_{\i+\hat y} P^y_i P^y_{i+\hat y} ]\nonumber \\
&-J_f \sum_{<i,j>} \vec S_i \cdot \vec S_j\nonumber \\
&+J_2 \sum_{<i,k>} \vec S_i \cdot \vec S_k.\nonumber\\
\end{eqnarray}
Here $P^x_i$ ($P^y_i$) are one if the $d_{xz}$ ($d_{xy}$) orbital is
occupied at $i$ and zero otherwise and $S_i$ are the spin operators
at site $i$. The sum over $<i,j>$ runs over nearest neighbors, and the
sum over $<i,k>$ runs over the second neighbors on the square lattice.

The ground state of the model is orbital ordered with $P^x_i=1$ 
and $P^y_i=0$ for all $i$ or vice versa. The spin system in this
orbital ordered state is unfrustrated and has a nearest neighbor
antiferromagnetic coupling $J_{1a}=J_1-J_f$ along $x$ and a ferromagnetic
coupling $J_{1b}=-J_f$ along $y$. Thus the system will have
Neel order. Linear spin-wave theory should be a good approximation
for the spectra of this model. Using Holstein-Primakoff transformations,
we obtain the spin-wave dispersion:
\begin{equation}
\omega_k= 4 S J_2 \sqrt{(A_k^2-B_k^2)}
\end{equation}
with
\begin{equation}
A_k=1+\alpha-\beta+\beta \cos{k_y},
\end{equation}
and,
\begin{equation}
B_k=\cos(k_x) (\cos(k_y)+\alpha).
\end{equation}
Here, $\alpha=J_{1a}/(2 J_2)$, and $\beta= J_{2a}/(2 J_2)$. The condition
for ($\pi,\pi$) point to be a maximum along $k_y$ is that $\alpha>1$ and
$\beta$ be negative and small. A sketch of the resulting spectra is
shown in Fig.~1.
Parametrically this is the model obtained
in the electronic structure calculations of Han et al.\cite{han} And, this is
exactly the form of the model that gives an excellent fit to the neutron
scattering spectra, where Zhao et al quote the values for CaFe$_2$As$_2$
($SJ_{1a}=49.9\pm 9.9$meV, 
$SJ_{1b}=-5.7\pm 4.5$meV, $SJ_2=18.9\pm 3.4$meV).\cite{zhao}

Because this model has a local Ising like degree of freedom associated
with orbital occupation, it should have a finite temperature phase
transition. However, the energetics of the Ising order are mediated by
ordering the spins. So, it is not clear that the universality class
of the transition must necessarily be that of the 2D Ising model.
There could be long-range effective interactions and that may lead
to a first order phase transition. In future work, such details can be investigated
numerically. Such an orbital order will clearly couple to lattice
distortions and a lack of tetragonal symmetry. 
Note, that unlike
the $J_1-J_2$ model, this Ising order parameter is much less subtle here
and the distortions should be more prominent. 
This is exactly what
one finds in the phenomenology of these materials. 

The model presented above gives a simple scenario for the observed
magnetism and structural transition in the pnictide materials. However,
it is certainly not a complete model for the system. At the very least,
this phenomena must coexist with other $d$ electrons, which are
itinerant and can contribute further to the magnetism. Electronic
structure studies with local spin-density approximation (LSDA) provide further
clues to the unusual physics. It is found that the local moment and
magnetism is highly sensitive to the assumed spin configuration. 
Not all choice of LSDA patterns lead to well formed moments.
The ($\pi,0$) pattern leads to particularly robust magnetism, whereas ferromagnetic
ordering leads to almost no moment at all.

We can view such a calculation as providing further information on
orbital occupations, when a spin-density is imposed on the system. 
Fixing the
($\pi,0$)-state in LSDA pattern drives the system to the orbital ordered
magnetic state discussed above. 
In contrast, fixing a ferromagnetic spin configuration, not only need not
cause orbital polarization, it fails to set the occupation of the
$d_{xz}$ and $d_{yz}$ orbitals at a level needed for moment formation.
This suggests that antiferromagnetic spin arrangement on neighboring
sites, and the resulting gain in the exchange energy $J_1$ is necessary
to stabilize the local moment. One way to view the situation is that
electrons in other, perhaps weakly correlated, itinerant orbitals act
as a reservoir for these strongly correlated orbitals. The gain in the
superexchange energy, when neighboring spins are anti-parallel, alters
their chemical potential in a way as to cause them to become Mott 
insulating, with occupancy close to unity and well formed moments.
This is an example of an orbital selective Mott transition, which
is driven by superexchange interaction between neighboring iron atoms.
It would be interesting to further study this by Dynamical Mean Field Theory (DMFT) and
its generalizations.\cite{georges}
The magnetic moment formed in these
orbitals, via the local Hund's coupling, 
can further polarize other $d$-electrons leading to an even enhanced moment as
found in the electronic structre calculations.


An intersting aspect of the model is that while the system may be
highly frustrated above the ordering temperature, the spins
become unfrustrated below the ordering temperature.
Yet, the ordered moment can be reduced significantly
by quantum fluctuations because we are dealing
effectively with a quasi-one dimensional spin-half system. 
Indeed, the electronic structure calculations show the bands to have much
larger dispersion along the direction of antiferromagnetic order.\cite{han}
$J_1$ being much larger than all other exchanges means that there will
be strong one-dimensional fluctuations and that will reduce the
ordered moment.
Any additional
moment in the system, coming from the polarization of other d-electrons through the
Hunds coupling,
will be subservient to this primary variable. Hence, many factors
can contribute to the full moment and indeed these systems show
a lot of variation from one material to another.

The most direct way to confirm such a scenario would be to 
look for orbital order through X-ray scattering. Since the pioneering
studies of Kugel and Khomskii,\cite{kugel} the subject of orbital
order has come a long way and is known to determine key aspects
of low energy physics of many materials. 
The electronic structure studies of these materials by
Han et al imply that even without any lattice distortion,
the tetragonal symmetry must be spontaneously broken. Orbital
order is a natural and perhaps only candidate for such a broken symmetry,
where the exchange constants themselves will be different along the
$X$ and $Y$ axes.
However, almost certainly, the orbital order is more subtle in these systems than
the simplified model suggests.
The occupation of $d_{xz}$ orbitals may exceed the occupation of
$d_{yz}$ orbitals by a small amount. Such a small symmetry breaking
may be sufficient to cause the exchange interactions to be antiferromagnetic
in one direction and ferromagnetic in the other. Nevertheless,
our model serves as a good zeroth order starting point for understanding this behavior.
It would be useful to look for such occupation numbers in more detail in the
electronic structure calculations.

The orbital degrees of freedom should couple strongly to the lattice.
Hence, they should have
long time scales associated with them and
may act like quenched impurities and give rise to glassy
behavior. It would be interesting to look for such effects in experiments.
Furthermore, the role of vacancies would be very different, in
such a system, from usual local moment antiferromagnets. Rather than form moments
around vacancies, one is more likely to destroy the Mott behavior
altogether.\cite{mazin}

To get a complete picture for this material, one needs to consider a
multi-band model with all the low energy electrons. 
However, this analysis shows that in some respects this system may be much closer
to the cuprates than previously thought and hiding inside the
multi-electron weakly correlated electronic system maybe a spin-half
strongly correlated component which drives the magnetic phenomena.
Whether this also plays a role in superconductivity remains to be
seen.

\begin{acknowledgements}
I would like to thank G. Uhrig, O Sushkov, S. Savrasov and W. Pickett
for useful discussions.
\end{acknowledgements}


\begin{thebibliography}{2}

\bibitem{kamihara} Y. Kamihara {\it et al}, J. Am. Chem. Soc. {\bf 130}, 3296 (2008).

\bibitem{cruz} C. de la Cruz et al, Nature 453, 899, (2008);
H. H. Klaus et al, PRL 101, 077005 (2008).

\bibitem{dong} J. Dong et al PRL 83, 27006 (2008).

\bibitem{yin} Z. P, Yin, {\it et al}, PRL 101, 047001 (2008).

\bibitem{cao} C. Cao, P. J. Hirschfeld, and H. P. Cheng, PRB 77, 220506 (2008).

\bibitem{ma} F. Ma and Z. Y. Lu, PRB 78, 033111 (2008).

\bibitem{yildrim} T. Yildrim, PRL 101, 057010 (2008).

\bibitem{wu} J. Wu, P. Phillips, A. H. C. Neto, PRL 101, 126401 (2008).

\bibitem{haule} K. Haule, J. H. Sjim and G. Kotliar, PRL 100, 226402 (2008).

\bibitem{anderson} P. W. Anderson, Science 235, 1196 (1987).

\bibitem{si} Q. Si and E. Abrahams, Phys. Rev. Lett. 101, 076401 (2008).

\bibitem{fang} C. Fang et al PRB 78, 052507 (2008).

\bibitem{xu} C. Xu, M. Mueller and S. Sachdev, PRB 78, 020501 (2008).

\bibitem{mazin} I. I. Mazin and M. D. Johannes, Nat. Phys. 5, 141 (2009).

\bibitem{raghu} S. Raghu {\it et al} PRB 77, 220503 (2008).

\bibitem{ran} Y. Ran et al, PRB 79, 014505 (2009).

\bibitem{chandra} P. Chandra, P. Coleman and A. I. Larkin, PRL 64, 88 (1990).

\bibitem{capriotti} L. Capriotti et al, PRL 92, 157202 (2004);
C. Weber et al PRL 91, 177202 (2003).

\bibitem{uhrig} G. Uhrig et al PRB to appear

\bibitem{zhao1} J. Zhao et al, PRL 101, 167203 (2008).

\bibitem{ewings} R. A. Ewings et al arXiv:0808.2836.

\bibitem{shender} E. F. Shender, Soviet Phys. JETP 56, 178 (1982).

\bibitem{singh} R. R. P. Singh et al PRL 91, 017201 (2003).

\bibitem{zhao}
J. Zhao et al, arXiv:0903.2686v1.

\bibitem{han}
M. J. Han et al Phys. Rev. Lett. 102, 107003 (2009).

\bibitem{georges} A. Georges {\it et al}, Rev. Mod. Phys. {\bf 68}, 13 (1996).

\bibitem{kugel}
K. I. Kugel and D. I. Khomskii, Sov. Phys. Usp. 136, 621 (1984).

\end{thebibliography}

\vskip 0.8in

\begin{figure}[!h]\label{Fig1}
\resizebox{85mm}{!}{\includegraphics{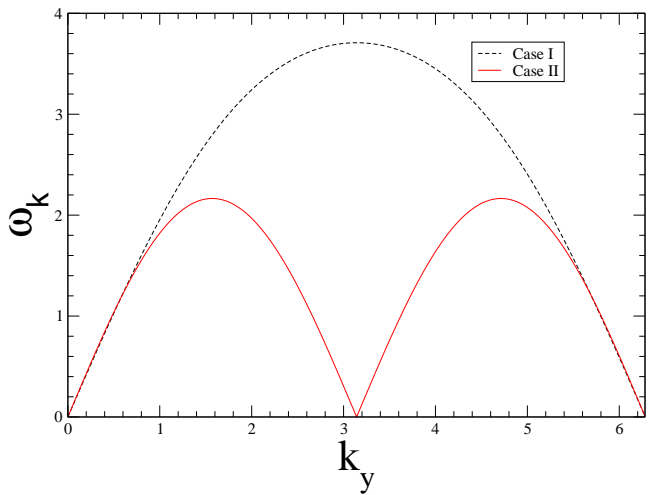}}
\caption{\label{fig:Fig1} Linear spin-wave spectra of the model through the Neel
point ($\pi,\pi$).
Case I refers to an antiferromagnetic coupling along x and a ferromagnetic coupling along y,
where as case II refers to equal antiferromagnetic coupling along both axes. 
In the latter case, order
by disorder phenomena opens a gap at ($q_y=\pi$) but it leaves the point
a mimima. }
\end{figure}

\end{document}